\documentclass[twocolumn,showpacs,aps,prl]{revtex4}
\usepackage{amsmath,amssymb}
\usepackage{array}
\usepackage [pdftex]{graphicx}
\usepackage{color}
\usepackage{natbib}
\usepackage[thinspace,thinqspace,squaren,textstyle]{SIunits}
\usepackage{subfigure}

\newcommand{\Matlab}{M{\scriptsize ATLAB}$^\circledR$~}

\definecolor{darkred}{rgb}{0.8,0.0,0.0}
\definecolor{lightblue}{rgb}{0.0,0.3,0.9}
\definecolor{darkred}{rgb}{0.8,0.0,0.0}

\begin{document}

\title{The Arnold-Tongue of Coupled Acoustic Oscillators}%

\author{Jost Fischer$^1$}
\email[corresponding author: ]{jost.fischer@uni-potsdam.de}
\author{Steffen Bergweiler$^2$}
\author{Markus, Abel$^{1,3,4}$}

\affiliation{$^1$Department of Physics and Astronomy, Potsdam University, Karl-Liebknecht-Str. 24, D-14476, Potsdam-Golm, Germany}
%\affiliation{Department of Physics and Astronomy, Potsdam University, Karl-Liebknecht-Str. 24, D-14476, Potsdam-Golm, Germany}
\affiliation{$^2$Eichbergstraße 6, D-96364 Marktrodach}
\affiliation{$^3$LEMTA - UMR 7563 (CNRS-INPL-UHP), 54504 Vandoevre, France}
\affiliation{$^4$Ambrosys GmbH, Potsdam, Germany}
\date{\today}%

\begin{abstract}

Wind-driven sound generation is a source of anger and pleasure,
depending on the situation: airframe and car noise, or combustion noise 
are some of the most disturbing environmental pollutions, whereas
musical instruments are sources of joy. We present an experiment on two 
coupled sound sources -organ pipes- together with a theoretical model which takes 
into account the underlying physics. Our focus is the Arnold tongue which 
quantitatively captures the interaction of the sound sources, we obtain
very good agreement of model and experiment, the results are supported
by very detailed CFD computations.
\end{abstract}

\maketitle

% 
% 43.25.+y 	Nonlinear acoustics
% 02.60.Gf 	Algorithms for functional approximation
%  05.45.-a 	Nonlinear dynamics and chaos
% 05.45.Gg 	Control of chaos, applications of chaos 
% 05.45.Tp 	Time series analysis 
% 05.45.Xt 	Synchronization; coupled oscillators 
% 07.05.Kf 	Data analysis: algorithms and implementation; data management
% 07.05.Tp 	Computer modeling and simulation 

\maketitle

Understanding wind-driven sound generation and -radiation is of high importance 
in many everyday situations, as noise from moving objects, whistling noise, combustion noise, 
industrial noise, or -to name a beautiful example- musical instruments.
Models, which describe such systems as self-sustained oscillators have been 
developed \cite{Abel-Ahnert-Bergweiler-09a,Abel-Bergweiler-Multhaupt-06}, and put these
systems in a very general perspective with the results having impact on 
synchronization community, including biosystems, lasers, mechanics, or social systems
\cite{Pikovsky-Rosenblum-Kurths-01}.
In a previous publication, an organ pipe, externally driven by a loudspeaker at a fixed position,
was investigated with focus on the detuning with the coupling varied by the speakers 
amplitude \cite{Abel-Ahnert-Bergweiler-09a}. However, for 
real aeroacoustical systems synchronization changes 
drastically with distance, or coupling, as we demonstrate hereafter. 
Here, we focus on the coupling mechanism by analyzing experimental
data in comparison with theoretical modeling. Results are supported by detailed
simulations of the compressible Navier-Stokes equations.
% As a perfect study subject, we use coupled organ pipes,
% and point out that the results are general and hold for all self-sustained oscillatory systems
% with sound generation and -radiation.
We present results on two coupled pipes, based on a refined experiment with detailed measurement of the phase relation. As in previous setups, we record the interfered signal at one microphone and analyze the 
synchronization regions and the measured Arnold tongue. 

In contrast to a simple model with direct coupling, we find an Arnold tongue which shows 
strongly nonlinear behavior. This is typical for systems with nontrivial coupling.
Consequently, we focus on the coupling mechanism:
we model the system with an empirical factor that considers
the coupling between the sound-generating wind field, and the wave propagation, which in turn
involves $1/r$ attenuation of the sound field and a delay term which reflects the time delayed
coupling of the field generated at the place of one pipe with the other one at a different place. 
The results of model and measurement coincide
very well in the far field, in the near field the sound propagation rather follows an $1/r^2$
law such that deviations occur naturally. 

Let us briefly sketch the typical functioning of an organ pipe. Energy is supplied steadily by
the wind system through the pipe foot and establishes a turbulent vortex
street. Each time a vortex detaches, a pressure fluctuation enters the
resonator, inside which characteristic waves are selected (resonator), and radiated at the
pipe mouth by an oscillating air-sheet \cite{Fabre-00,Fabre-Hirschberg-96} (oscillator). 
Inside the resonator energy is dissipated. The coupling of an external acoustical
field can be described by a (nonlinear) acoustical admittance \cite{Thwaites-Fletcher-83}. 
As argumented in \cite{Abel-Ahnert-Bergweiler-09a}, the air sheet is the source of
sound radiation, and the measurement at the microphone can be taken as the state
of the oscillator.

\begin{figure}[!thb]
%\begin{center}
\subfigure[]
    {
\includegraphics[draft=false,width=0.296\textwidth]{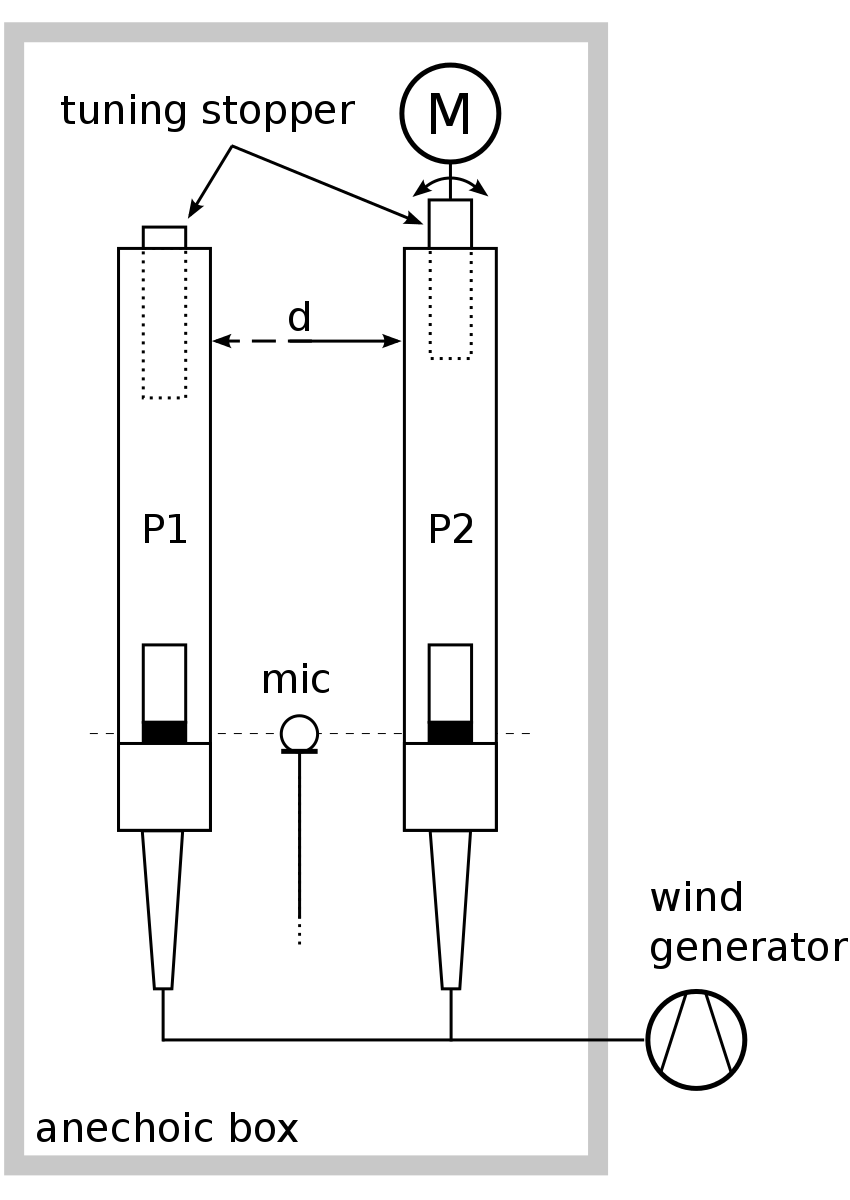}
}
\subfigure[]
    {
\includegraphics[draft=false,width=0.152\textwidth]{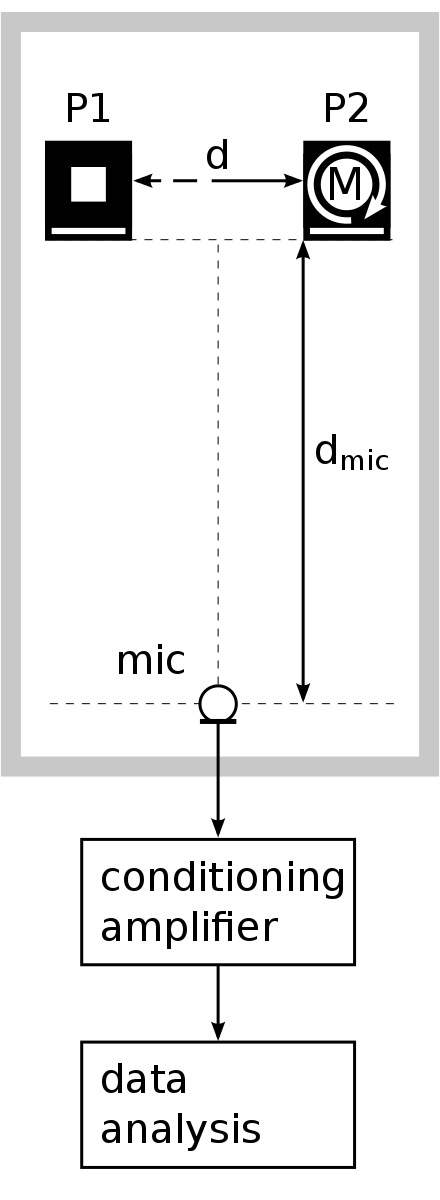}
}

  \caption{Experimental setup. (a) front view: P1, and P2: tunable organ
    pipes. Both are connected by flexible tubes to a wind generator,
    positioned outside the anechoic box, sketched as the thick  gray
    frame. The identically constructed, stopped pipes were mounted on
    a common bar, their distance d could be varied. The frequency
    $f_2$ of $P2$ was tuned by the step motor (M), 
    in steps of   $0.5\usk\hertz$ erhöht. (b) top view: the microphone
    was positioned at a distance $d_{mic}=1\usk\meter$ in  the
    midplane of P1 and P2. Its signal was amplified by a Br\"uel \& K\ae r Nexus 2690 amplifter and stored for
    later data analysis.}
  \label{fig:experimental_setup}
%\end{center}
\end{figure}

The setup was analogous to the one described in  \cite{Abel-Bergweiler-Multhaupt-06},
however with smaller, and thus higher-pitched, pipes.
The measurements were carried out on a miniature organ especially made
by Alexander Schuke GmbH \cite{Schuke}.
The  air-supply was connected via a mechanical regulating-valve 
to a wind-belt, the two pipes were 
joined  directly to the wind-belt by flexible tubes. 
Whereas in  \cite{Abel-Bergweiler-Multhaupt-06}, the pipes stood
side-by-side, here, both were mounted on a horizontal bar, along
which their position, i.e. their mutual distance, was controlled,
cf. Fig.~\ref{fig:experimental_setup}.  Another difference was the
size: here, we used smaller, stopped pipes, tuned at $720\usk\hertz$, with 
a more suitable wavelength $\lambda\simeq 0.48\usk\metre$ for distance variation, details
on the pipe geometry are given in the supplement.

This allowed us to use of the anechoic chamber of Potsdam University 
($B\cdot H \cdot T=1.6\usk\metre \cdot 2.0\usk\metre \cdot
1.3\usk\metre$). During the experiment humidity, temperature, and
pressure were monitored and held constant at the normal conditions
$L_h=20\usk$\%, $T_0=20\usk\celsius$, $p_0=1013.25\usk\hecto\pascal$.

Our main goal was the investigation of the Arnold-tongue, i.e. a scan
of the parameter space coupling, $\epsilon$, vs. detuning , $\Delta f$.
For the coupling strength, we mounted the pipes $P1$, and $P2$
 on a common horizontal bar, along
which  their mutual distance $d$, was
controlled (Fig.~\ref{fig:experimental_setup}). 
%This way, one of the fundamental synchronization parameters, the coupling
%strength $\epsilon$, could be varied. 
To vary the  second important
parameter, the frequency detuning $\Delta f$,  $P1$ was tuned at
$f_{P1}=720\usk\hertz$, while $f_{P2}$ was varied by a step motor in the
range of $700 - 740\usk\hertz$, and  $680 - 760\usk\hertz$ for large
and small distances, respectively (small distances
correspond to large coupling and \textit{vice versa}). 
The pipes were well-tuned independent on eachother before coupling them.

To explore parameter space, the pipes were positioned at the distances 
$1,10, 30, 50, 75, 100, 200, 300$, and $400\usk\milli\metre$. 
For each distance the frequency was varied as 
described above in steps of $0.5\usk\hertz$, for details see \cite{supplement}.  
%Measurement time was $8\usk\second$ for each run, such that the total time amounts
% to approximately $2\usk\hour$, which is comparatively fast.
The acoustic signal was measured by a Br\"uel \& K\ae r 4191 condenser microphone
at $d_{mic}=1\usk\metre$ distance, from the pipes midpoint, 
at the centerline between the sound sources cf. Fig.~\ref{fig:experimental_setup}, which satisfies the farfield condition $d > \lambda$.
The sampling rate was  $f_s=44.1\usk\kilo\hertz$ with a resolution of
$16\usk$bit. The data analysis and signal processing was programmed
using   \Matlab.\\

\begin{figure*}[!htb]
\begin{center}
\subfigure
    {
\includegraphics[draft=false,width=1.0\textwidth]{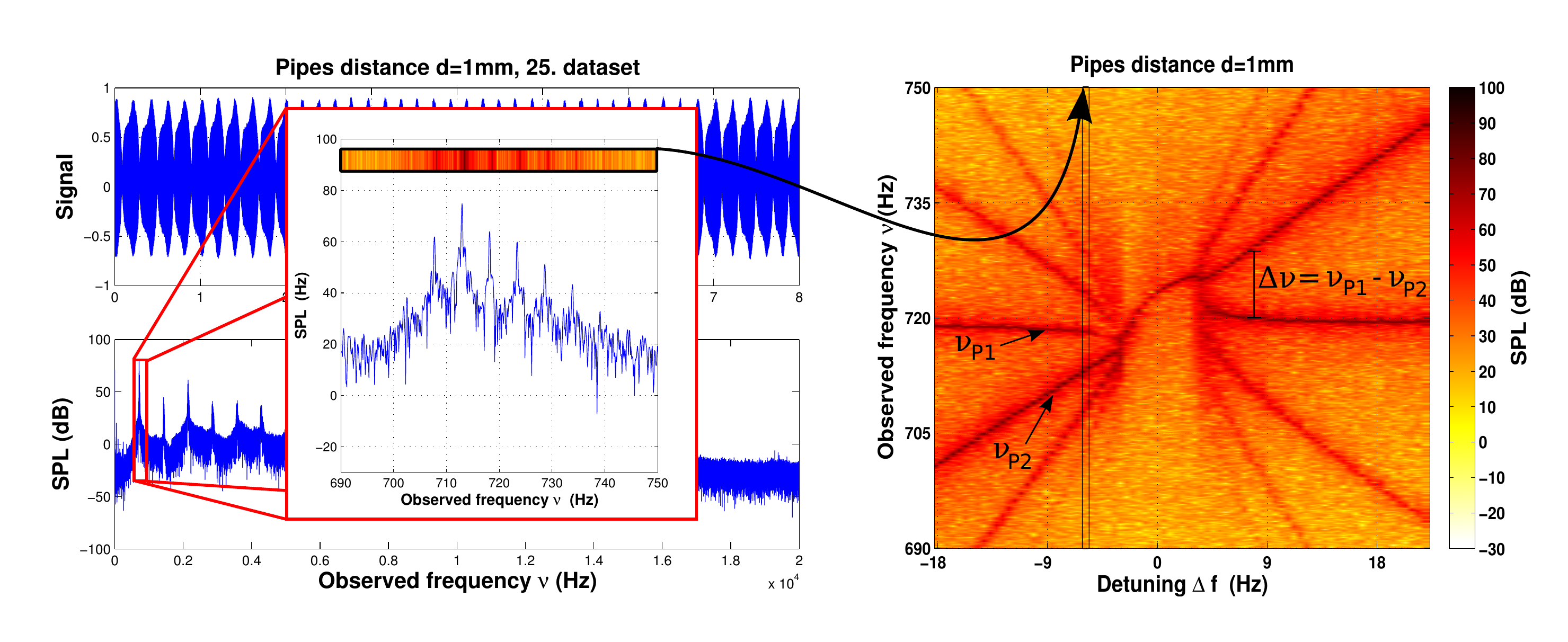}
}
  \caption{From signal to synchronization plot: Left: the signal is Fourier-transformed, 
the resulting spectrum is color coded, and the maxima are determined. Right: For each 
detuning $\Delta f$ we obtain such a bar; further we combine such bars for
each distance $d$ and obtain the colorful plot on the right, here for $d=1\usk\milli\metre$. 
The maxima give the measured frequency difference $\Delta \nu$, which vanishes for synchronized
behavior. The amplitude of the measured signal achieves maximum for in-phase and
minimum for anti-phase synchronization.}
\label{fig:Procedure_sound_signal_to_CCP}
\end{center}
\end{figure*}

\begin{figure*} [!htb]
\begin{center}

\subfigure[]
{
  \includegraphics[draft=false,width=0.30\textwidth ]{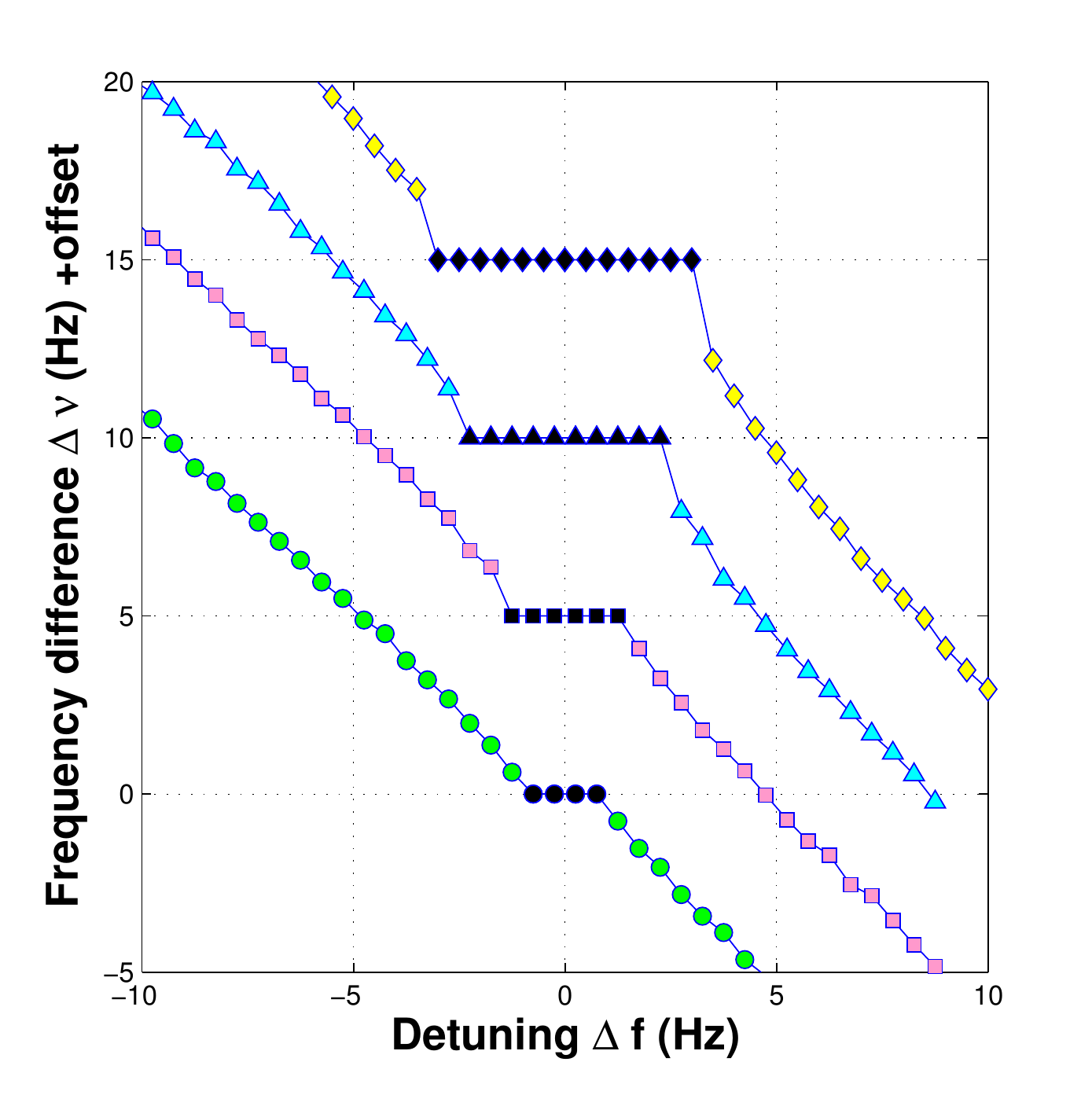}
\label{fig:Synchro_plateaus_1mm_10mm_50mm_100mm}
  }
\subfigure[]
{
  \includegraphics[draft=false,width=0.30\textwidth ]{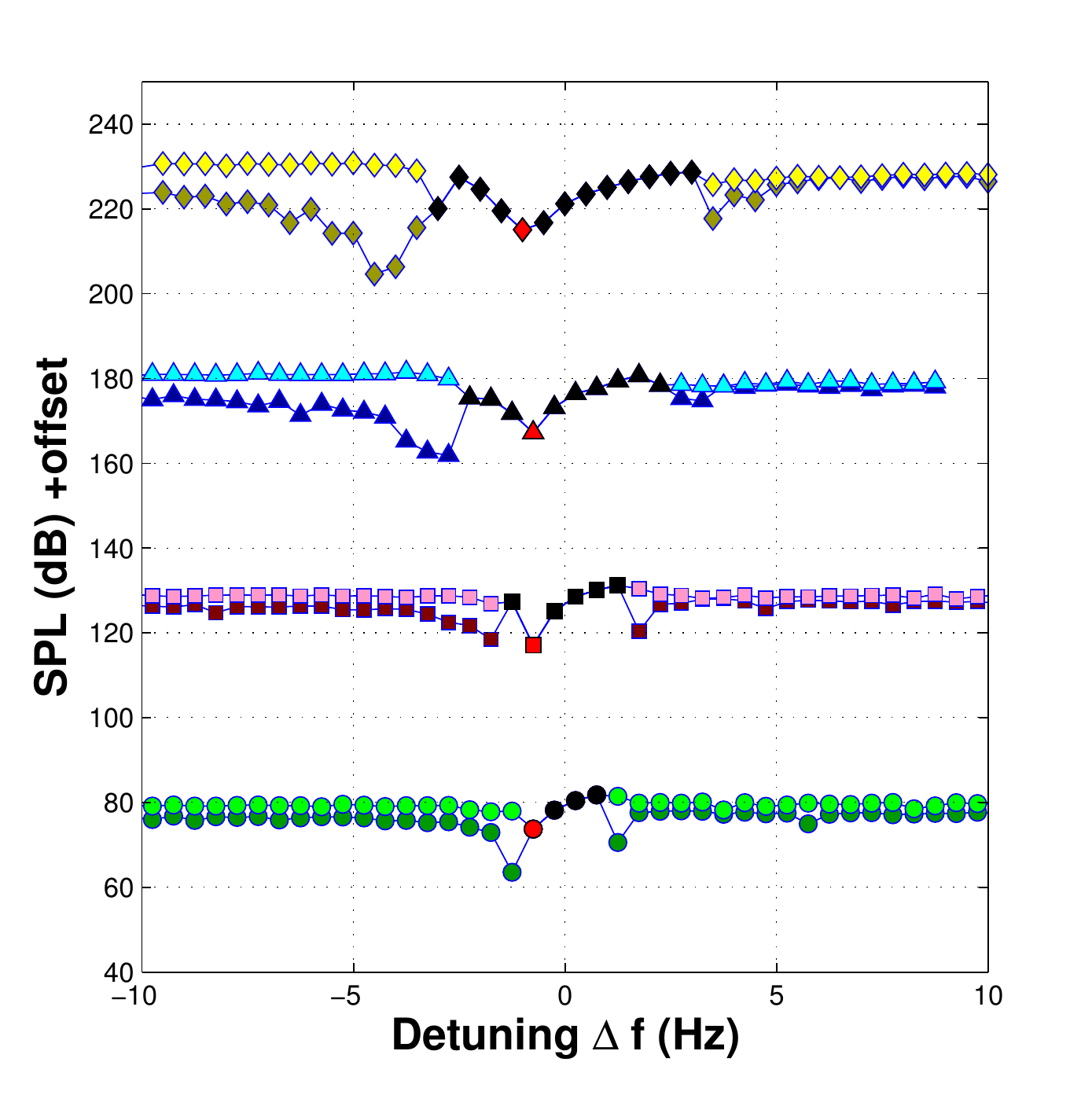}
\label{fig:Amplitudes_1mm_10mm_50mm_100mm}
  }
\subfigure[]
{
\includegraphics[draft=false,width=0.30\textwidth ]{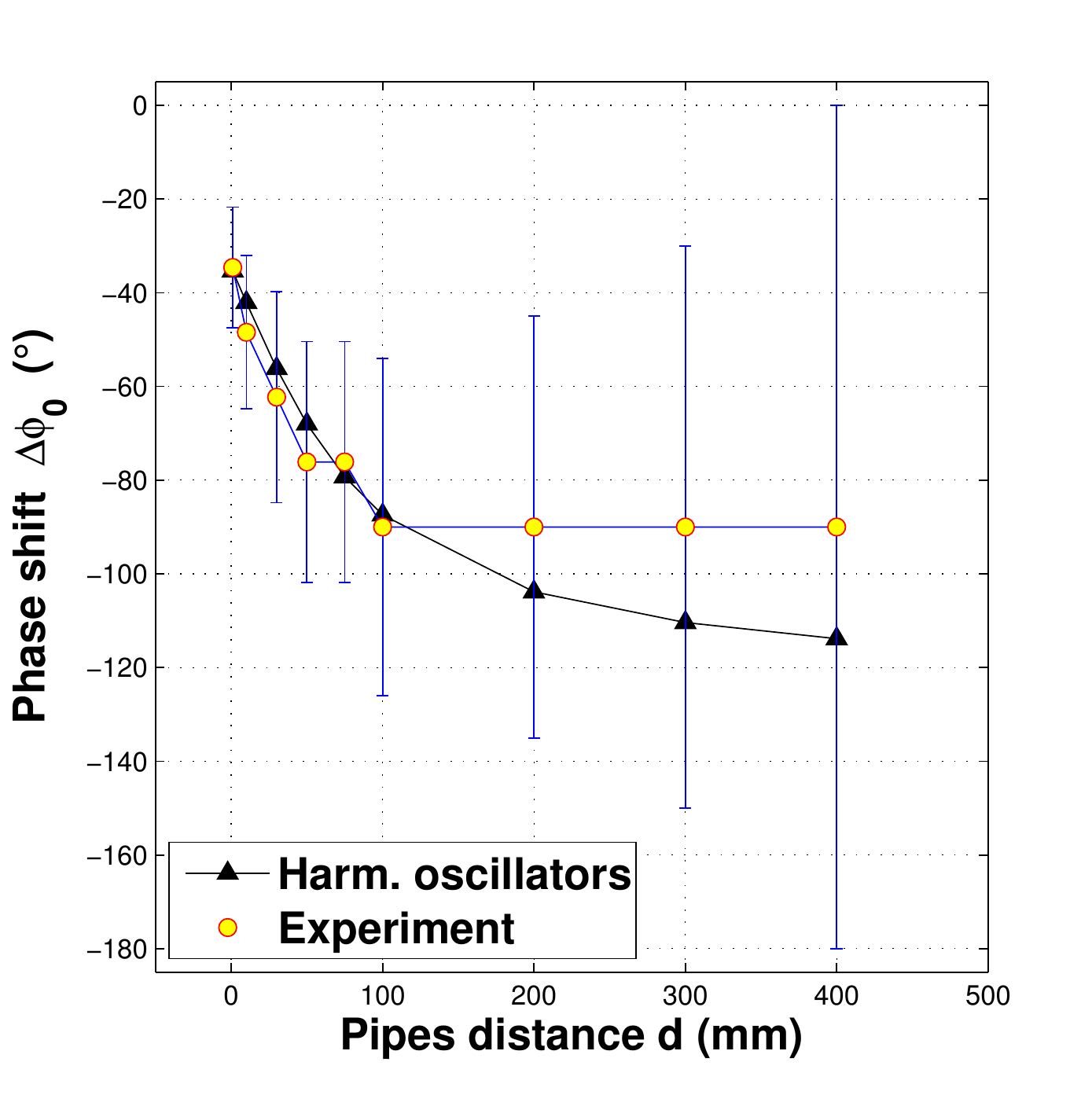}
\label{fig:DeltaPhi}
}
  \caption{Left:Synchronisation plateau, for good  for 4 exemplary distances $d=100, 50, 10$ and $1\usk\milli\metre$. The 
y-axis is plotted for $d=100\usk\milli\metre$. For the distances $d=50,10$ and 
$1\usk\milli\metre$ an offset of $5, 10$ and $15\usk\hertz$ is added to separate 
the curves. 
  All nine figures are found in \cite{supplement}. 
%  {\bf \color{red} Legende raus, stattdessen in die caption. }. 
  Right: Analogous plot for the peak amplitudes of the two pipes. 
  The y-axis is plotted for $d=100\usk\milli\metre$. 
  For the distances $d=50,10$ and $1\usk\milli\metre$ an offset of $50, 100$ and 
$150\usk\hertz$ is added to separate the curves. The synchronisation regions are 
labeled black. One recognizes a shift of the relative phases $\Delta \Phi_0$ 
while the
  distance is increased. The signal recorded at a microphone in the far field 
coincides nicely with the signal one would obtain for two harmonically oscillating sources at the position
of the two pipes. The vertical bars denote the errors which follow from the frequency discretization.}
  \label{fig:Synchro_Plateaus}
\end{center}
\end{figure*}

% \subsection{Data Analysis}
Data Analysis: A synchronization plot is generated by the procedure illustrated in 
Fig.~\ref{fig:Procedure_sound_signal_to_CCP}:
We are interested in the freqency difference and amplitude 
measured at the microphone,
since this gives us information on the relative phases $\phi_{01},\phi_{02}$ of 
the two pipes, which in turn yield negative or positive interference, 
causing a mutual cancellation or amplification 
\cite{Stanzial-00,Angster-93,Abel-Bergweiler-Multhaupt-06,Abel-Ahnert-Bergweiler-09a}.
% for a single measurement at distance $d$ and frequency detuning $\Delta 
% f=f_1-f_2$.
% The \textit{measured} frequency difference is denoted by $\Delta \nu$ for 
% convenience.
% By Fourier transform  the signal we compute the spectrum of the sound pressure level 
% $\mathrm{SPL} = 20 log_{10} \frac{p}{p_0}$, with $p_0=1013.25\usk\hecto\pascal$)
% for each measurement with $(d,\Delta f)$. 
% The amplitude is displayed by a color-coded SPL.
To obtain a synchronization plot $\Delta \nu$ vs. $\Delta f$, we 
first Fourier transform the signal
(Fig.~\ref{fig:Procedure_sound_signal_to_CCP}, left),  and 
consider the region $690\usk\hertz \leq \nu_{P1,P2} \leq 750\usk\hertz$. 
This part of the spectrum is turned into color-coded SPL
stripe.  We collect these stripes for each detuning 
and obtain a spectral plot $\nu_{meas}$ vs. $\Delta f$, with either 
several or one vertical frequencies corresponding to 
(non)synchronized behavior. The amplitude allows to infer the phase relation of
the two pipes, cf. Fig.~\ref{fig:Procedure_sound_signal_to_CCP}, (right),
The final synchronization plot is obtained by identifying the two main peaks 
$\nu_1$ and $\nu_2$ which yield $\Delta \nu$. 
This procedure is repeated for every distance, the resulting curves are plotted in 
Fig.~\ref{fig:Synchro_Plateaus}. 

In addition to the phases, we plot the SPL at the microphone (Fig.~\ref{fig:Synchro_Plateaus}, b)
From its minimum we obtain the phase difference between the two pipes. Within the error 
bars, the plot confirms the idea that the interference is as for two harmonic oscillators 
with phase difrerence $\Delta \Phi_0$ at the microphone.
% The detection of the transition to synchronization is
% thereby given by the frequency resolution of $0.5$Hz.

\begin{figure*} [!htb]
\begin{center}
%\subfigure[]
%    {
\includegraphics[draft=false,width=0.45\textwidth]{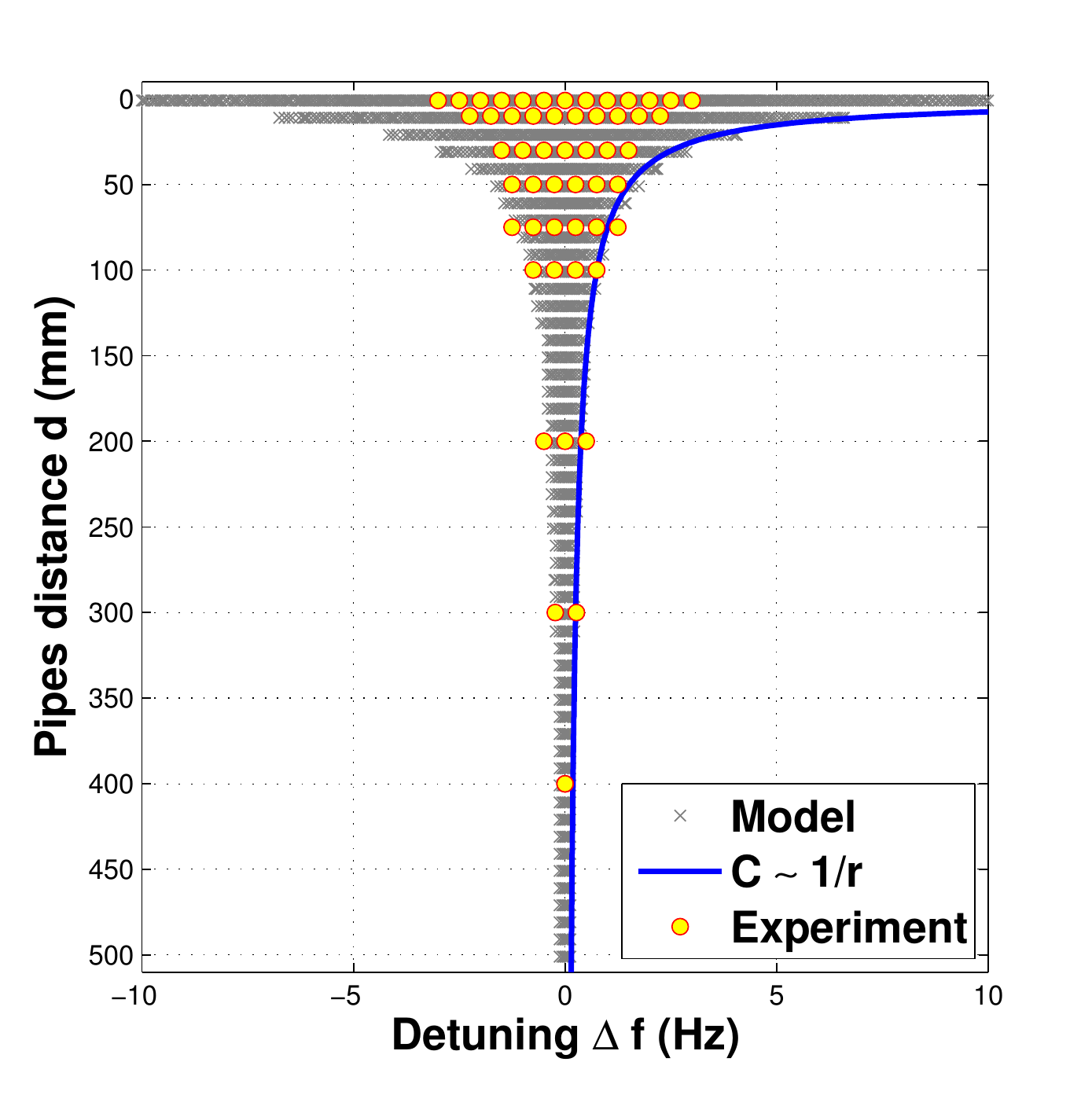}
 \label{fig:Arnold_tongue_and_Model}
%}
\caption{Arnold tongue for two coupled organ pipes. Circles: The horizontal error bars correspond
to the frequency resolution which sets the accuracy for the detection of the transition 
to synchronization with changing detuning. The stars correspond to results for the model, given in
Eq.\ref{eq:model}. The ordinate is plotted top-down, because the coupling decreases with 
increasing distance. The straight line corresponds to the 
expected coupling decrease with $1/d$. Experiment and model agree very well.}
\label{fig:arnoldtongue}
\end{center}
\end{figure*}

Given one such curve for each distance, we retrieve the Arnold tongue in $(d,\Delta f)$-space. 
We recognize an Arnold tongue with strongly curved boundaries (Fig.~\ref{fig:arnoldtongue}), typical for real experiments 
with delay and complicated couplings \cite{Flunkert-Fischer-Schoell-13}. 
In order to understand this behaviour from physical reasoning, we now develop a model consistent with 
aeroacoustics and nonlinear dynamics.

In order to understand the observed behaviour we have to recall that our observed system 
obeys the compressible Navier-Stokes-Equation with suitable boundary conditions. 
The signal measured is the acoustic field (not the aerodynamic wind field), 
at the microphone position. In order to reproduce the measured curve, 
we have to 
i) find a good model for a single organ pipe as a self-excited oscillator, 
ii) find a model for the emission of acoustical waves from an oscillating jet, 
iii) describe the propagation from pipe to pipe
iv) find a model for the coupling of the acoustic field - the one emitted by the partner pipe- 
to the jet of a pipe.
In the following we describe the situation that sound emitted at $x_2$ by pipe 2 influences 
the jet at $x_1$, pipe 1. Further, we understand the locations in a coarse-grained sense, i.e. the whole
jet is located at $x_1$, defined appropriately, e.g. as the mean jet position.

i) Single organ pipe. A self-sustained
oscillator consists of an oscillating unit, and energy source and sink, 
both possibly nonlinear. In our case, the oscillating unit is the jet, which exits 
from the pipe mouth and can be described by its displacement normal to the pipe longer axis.
Its frequency is set by the resonator: initially, the jet impinges on the
labium and a turbulent vortex street develops. 

Each time a vortex detaches a pressure wave travels upwards inside the 
resonator, is reflected at the closed end and travels back to hit the labium after 
the period $T=c/(\lambda/4)$ with $T$, with $c$ the speed of sound and $\lambda$ the wavelength.
The high pressure in turn triggers the detachment of the next vortex,
and very quickly a regular oscillation is established \cite{Howe-75,Fabre-Hirschberg-96}. 
Of course, this is an idealized description
and the true 3D pipe shows some quite complicated additional effects; the main physics, however,
is covered well by this picture; this is supported by many numerical runs, cf. \cite{supplement}.
% We would like to point out that we consider the jet as oscillator, because the sound 
% is mainly radiated by the velocity fluctuations at the jet and its interaction with 
% the labium \cite{Howe,Hirschberg}. 
The energy is supplied by the pressure difference at the jet outlet, such 
that the oscillator carries its own power supply as the mean jet velocity.
To first order the period depends linearly on the jet velocity.
The energy sink is here twofold: nonlinear energy dissipation inside the 
resonator and at the walls, and  radiation of an acoustical wave.
From \cite{Abel-Ahnert-Bergweiler-09a} we know that empirically one finds 
an excellent match for a van-der-Pol oscillator with additional weak higher-order 
nonlinear damping.

ii) The jet at $x_2$ emits a (sound) velocity wave, which is related to pressure by
$p_2=Z\,v\,A_{jet}$ with $Z$ the acoustical impedance at the jet. This is very hard to determine
from first principles. Since the jet is an extended source where each point emits
sound, a coarse-grained description must involve spatio-temporal averaging over 
the source region and long enough time, this is subject of future numerical work. Here,
we assume no phase shift and direct proportionality 
of emitted sound and jet displacement with an empirical factor $C_{em}$.

iii) A spherical (pressure) wave, emitted at $x_2$ with amplitude $p_2$ and frequency $f_2$ 
propagates to  $x_1$ according to 
$p_2(x_1,t_1)=\frac{p_2(x_2,t_2)}{d}\cdot  e^{i(k_2 d - \omega_2 \tau)}$,
with $d=x_1-x_2$, and $\tau=t_1-t_2$ \cite{Ingard-08}. 

iv) The integral force on the jet region at $x_1$ is $F_{in}=C_{in} p_{2}(x_1,t_1) \cdot A_{jet}$, with
$p_2$ the sound emitted by pipe 2, and $A_{jet}$ the area covered by the jet.

Alltogether, we obtain for the force from pipe 2 on jet 1: 
$F_{21}= C_{in}  \frac{p_2(x_2,t_2)}{d}\cdot  e^{i(k_2 d - \omega_2 \tau)} C_{em}$.
As usual, there is a difference between near field and far field in that pressure and
velocity show the characteristic phase difference of $\pi/2$ or 0 and $1/d$, or $1/d^2$, respectively.

For a complete model, we combine i)-iv), to obtain a model for the displacement
of the jet, $\xi$:
 \begin{eqnarray}
  \ddot{\xi}_1 - \mu_1 (1 - \beta_1 \xi_1^2)\dot{\xi}_1 +\omega_{01}^2 \xi_1  
  &=&   C_1\dot{\xi}_2  \\
  \ddot{\xi}_2 - \mu_2 (1 - \beta_2 \xi_2^2)\dot{\xi}_2 +\omega_{02}^2 \xi_2  
  &=&   C_2\dot{\xi}_1  \, ,
  \label{eq:model}
\end{eqnarray}
with $\mu_i$ the energy supply by velocity $\dot{\xi}_i$,  $\beta_i$ 
the nonlinear damping, $\omega_{0i}$ the individual angular frequency, 
these terms characterize the individual pipes ($i=1,2$). The coupling is modeled by 
the coefficients $C_i$ with $C_i=C_{i,em}\cdot C_{i,trans} \cdot C_{i,in}$.
where $C_1$, $C_2$ model the sound-fluid interaction. A more realistic model needs to include the
frequency shift during wave propagation, this is subject of ongoing research.

How does the coupling influence the Arnold tongue? We integrated Eq.~\ref{eq:model} numerically
using odeint \cite{odeint}. We do assume that both pipes stay at a fixed phase relation
after some transients, such that we neglect the 
oscillation $e^{i\omega t}$, and the relative phase change during propagation is captured by
the term $e^{-ik}$.  As a result, we obtain an Arnold tonge which 
coincides very well with the experimental data, cf. Fig.~\ref{fig:arnoldtongue}.
At a closer look, one finds deviations for high coupling (small distance).
This is explained by the fact that in the acoustic near field the $1/r$ law does not hold,
and rather a $1/r^2$ behavior is expected. Since the range of scales is too small,
one cannot clearly make out a power-law, cf.\cite{supplement}.
% 
% \paragraph{Discussion and Conclusion}

Summary: We investigated an improved experimental setup of two coupled wind-driven 
sound sources. As well controlled realization, we used organ pipes. 
The model for a single pipe presented in 
\cite{Abel-Ahnert-Bergweiler-09a} is consistent with pipes of various 
dimensions, indicating that our results can be transferred to other wind-driven oscillatory systems. 
Furthermore, we developed a model for the coupling of two
sound sources (two jets) by modeling the coupling sound-jet with an empirical factor,
the sound wave is modeled as a monopole, whose propagator includes attenuation with inverse distance.
The coupling into the jet is again modeled by a phase shift and an empirical factor.

The Arnold tongue measured shows a clear curvature which can be explained by
the above model in very good coincidence with the experiment in the far field. This has been
validated by numerical integration of the two coupled ODEs. 
In the near field, the sound field is highly angle-dependent and in general 
decays with squared inverse distance. This is again confirmed by the comparison 
of numerical and experimental results.

We see our results in a much more general context than musical acoustics:
on one hand, we have investigated one aspect of the fundamental question of sound emission by
a turbulent jet. On the other hand we have demonstrated how powerful the concept of 
synchronization can be applied even for 3D systems with turbulent behavior.
Indeed, we are running an experiment on coupled Rijke tubes, which are an excellent 
example for flame-induced combustion noise.
Eventually, we have contributed to improve the understanding of one of the most 
beautiful instruments humans have created - the organ.

\paragraph*{Acknowledgments.}
We acknowledge inspiring discussions with helpful remarks with
A. Pikovsky and M. Rosenblum. 
We thank  Alexander Schuke Potsdam Orgelbau GmbH for their active help
in pipe and wind supply construction and R. Gerhard for his great enthousiasm 
and constant support, including the use of the anechoic chamber 
needed for the measurement.
J. Fischer was supported by ZIM, grant ``Synchronization in Organ Pipes''

\bibliographystyle{unsrtnat}
%\bibliography{bibtex/organ,bibtex/mypeerrev,bibtex/articles}

\end{document}